\documentclass[11pt]{article}\usepackage[]{graphicx}\usepackage[]{color}
\makeatletter
\def\maxwidth{ %
  \ifdim\Gin@nat@width>\linewidth
    \linewidth
  \else
    \Gin@nat@width
  \fi
}
\makeatother

\usepackage{alltt}
\usepackage{bm}
\usepackage{geometry}
\usepackage{url}
\usepackage[american]{babel}
\usepackage{graphicx}
\usepackage{rotating}
\usepackage{tabularx}
\usepackage{amssymb,amsmath}
\usepackage{ifxetex,ifluatex}
\usepackage{fixltx2e} 

\usepackage[round,authoryear]{natbib}
\usepackage{tipa}

\IfFileExists{upquote.sty}{\usepackage{upquote}}{}
\ifnum 0\ifxetex 1\fi\ifluatex 1\fi=0 
  \usepackage[utf8]{inputenc}
\else 
  \ifxetex
    \usepackage{mathspec}
    \usepackage{xltxtra,xunicode}
  \else
    \usepackage{fontspec}
  \fi
  \defaultfontfeatures{Mapping=tex-text,Scale=MatchLowercase}
  
\fi

\IfFileExists{upquote.sty}{\usepackage{upquote}}{}

\title{
\rule[0.4cm]{\textwidth}{2pt}
{\bf Corrections for multiple comparisons in voxel-based lesion-symptom mapping}
\rule{\textwidth}{2pt} 
}

\begin{document}
\maketitle
\begin{center}
Daniel Mirman$^{a,b,}$\footnote{Address correspondence to Daniel Mirman, dan@danmirman.org. Portions of this work were previously described in	arXiv:1606.00475 [stat.AP].} \\
Jon-Frederick Landrigan$^c$ \\
Spiro Kokolis$^c$  \\
Sean Verillo$^c$ \\
Casey Ferrara$^b$ \\
Dorian Pustina$^d$ \\ \ \\
$^a$ University of Alabama at Birmingham, Birmingham, AL, USA\\
$^b$ Moss Rehabilitation Research Institute, Elkins Park, PA, USA \\
$^c$ Drexel University, Philadelphia, PA, USA\\
$^d$ University of Pennsylvania, Philadelphia, PA, USA\\
\end{center}

\begin{center}
  {\bf Abstract}
\end{center}

\noindent Voxel-based lesion-symptom mapping (VLSM) is an important method for basic and translational human neuroscience research. VLSM leverages modern neuroimaging analysis techniques to build on the classic approach of examining the relationship between location of brain damage and cognitive deficits. Testing an association between deficit severity and lesion status in each voxel involves very many individual tests and requires statistical correction for multiple comparisons. Several strategies have been adapted from analysis of functional neuroimaging data, though VLSM faces a more difficult trade-off between avoiding false positives and statistical power (missing true effects). Non-parametric, permutation-based methods are generally preferable because they do not make assumptions that are likely to be violated by skewed distributions of behavioral deficit (symptom) scores and by the necessary spatial contiguity of stroke lesions. We used simulated and real deficit scores from a sample of approximately 100 individuals with left hemisphere stroke to evaluate two such permutation-based approaches. Using permutation to set a minimum cluster size identified a region that systematically extended well beyond the true region, even under the most conservative settings tested here, making it ill-suited to identifying brain-behavior relationships. In contrast, generalizing the standard permutation-based family-wise error correction approach provided a principled way to balance false positives and false negatives. An implementation of this continuous permutation-based FWER correction method is available at https://gist.github.com/dmirman/05a92e0e9e0027f6fe6e528c648143d7\\


\section{Introduction}

Identifying relationships between location of brain damage and cognitive deficits is a foundational method in cognitive neuroscience, tracing its history at least to the behavioral neurologists of the mid-19th century (e.g., \citealp{Lichtheim1885}). Those early studies were based on individual case studies and, as data accumulated, researchers used lesion overlays to identify the locations where damage consistently produced deficits of interest. Recent advances in neuroimaging technology have allowed much finer-grained analyses at the level of individual voxels \citep{Bates2003, Rorden2004}. In voxel-based lesion-symptom mapping (VLSM), an association between deficit severity and lesion status (lesioned vs. not lesioned) is tested in each voxel, producing a statistical map of the strength of relationship between lesion status and deficit. However, this map is the result of individual tests across tens or even hundreds of thousands of voxels. 

The large number of tests involved in analysis of neuroimaging data requires some kind of statistical correction for multiple comparisons. Several strategies have been proposed, often by adaptation from analysis of functional neuroimaging (e.g., fMRI). The general approach of permutation-based correction for multiple comparisons can be implemented in many different ways, depending on what aspect of the results is to be controlled. Standard null hypothesis statistical tests compare the observed test statistic against a null distribution of test statistics to determine the likelihood of observing the result if the null hypothesis were true. Permutation-based statistical methods build the null distribution by permuting the real data, which has the important advantage of not making assumptions about the distributions of scores or test statistics -- assumptions that are likely to be violated by skewed distributions of behavioral deficit (symptom) scores and by the necessary spatial contiguity of stroke lesions. Building a null distribution based on permutations of real data offers a rather literal way to compute $p$-values: the $p$-value is literally the probability of observing a particular outcome if there were no relationship between the behavioral scores and lesion patterns (i.e., random permutations). A null distribution of the test statistic can built based on permutations of real data and used to reject voxels where the true analysis does not sufficiently differ (e.g., $p > 0.01$) from the permutation-based null distribution to warrant rejecting the null hypothesis (e.g., \citealp{Kimberg2007}). One standard strategy of correction for multple comparisons is to control voxel-level family-wise error rate (FWER), which is the probability of making one or more false positive (Type 1) errors among the entire set of tests. Controlling the probability of making one or more false positive errors is based on the idea that each test is critically related to the researcher's interpretation or inference, thus, a single false positive could potentially undermine the inference and needs to be controlled. This does not align with VLSM interpretation, which never depends on a single voxel. The misalignment between standard FWER and VLSM interpretation makes standard voxel-level FWER correction unnecessarily conservative: if no inferences are made based on a single voxel, then a single false positive voxel cannot be responsible for an invalid inference about lesion-symptom relations. 

An alternative correction approach is to focus on controlling errors at the level of the overall pattern of lesion-symptom relations while allowing one or more individual voxels to be false positives. One such strategy is to use permutations to determine a minimum cluster size. Setting a minimum cluster size is a common “clean-up” step in neuroimaging data analysis; the addition of a principled strategy for selecting the minimum cluster size is the critical component that turns this into a statistical correction method. Another such technique is False Discovery Rate (FDR), which quantifies the proportion of above-threshold results that can be expected to be false positives \citep{Genovese2002}. That is, at FDR threshold $q = 0.05$, 5\% of above-threshold voxels are expected to be false positives, which is likely to be substantially more than one voxel but not likely to affect interpretation of the overall patten (for a clear description see \citealp{Bennett2009}). FDR is widely used for analysis of functional neuroimaging data and VLSM; however, we have encountered informal criticism that FDR is inappropriate for VLSM, though we are not aware of any published analysis supporting such criticism. 

Correction for multiple comparisons is an attempt to manage variability, but it cannot remove all of the noise and leave all of the signal. Either some noise will get left behind or some of the signal will be removed. That is, there is an inherent trade-off between false positives and false negatives, incorrectly generalizing a result and overlooking a generalization that is warranted. By convention, data analysis requires setting a threshold to identify results that warrant rejection of the null hypothesis. There is a substantial price associated with adopting the conservative position that the probability of even a single false positive voxel needs to be controlled: VLSM analysis is based on a single data point per participant (each participant only has one lesion and only one deficit profile) and sample sizes are often limited by the practical challenges of recruiting and testing large numbers of participants with the targeted neurogenic deficits. This price is further exacerbated by publication bias: studies that meet the statistical threshold may be published, but studies that fall short are relegated to the “file drawer”, leaving a biased scientific literature. Publication bias also encourages various forms of “p-hacking” or “researcher degrees of freedom”, in which researchers try alternative analysis strategies (excluding certain “outlier” participants, transforming scores, etc.) until they find one that surpasses the statistical threshold. The result is a report that appears to use rigorous statistical methods, but the actual rate of false positives far exceeds the nominal $p$-value (e.g., \citealp{Simmons2011, Nosek2012, Gelman2014}). In addition to statistical soundness, the analytical strategy should allow researchers to transparently report their observations and the strength of the evidence that supports their conclusions.

The present study investigated two permutation-based methods of correcting for multiple comparisons in VLSM. The next section describes our investigation of using permutations to determine a minimum cluster size. Our analyses found that this approach produces consistent spill-over into neighboring regions (i.e., the identified region extends well beyong the boundaries of the true lesion-symptom relation), making it not very well-suited to identifying brain-behavior relationships. The subsequent section describes a generalization of the permutation-based FWER correction approach that captures some of the inferential advantages of FDR without making parameteric assumptions about the data. This approach makes it possible to balance control of false postives against risk false negatives, and to transparently report results in a way that allows others to evaluate the evidence. We also compare this approach with the parametric FDR method and describe conditions under which FDR may produce misleading results. The final section of this report summarizes our findings and conclusions, and discusses future directions.

\section{Minimum Cluster Size}

Using permutations to determine a minimum cluster size proceeds as follows: (1) permute behavioral data and conduct VLSM analysis, (2) apply a pre-set threshold for each voxel (e.g., $p < 0.0001$), (3) compute sizes of supra-threshold voxel clusters, (4) repeat steps 1-3 many times to build up a null distribution of supra-threshold cluster sizes (e.g., \citealp{Nichols2002}). This null distribution is the distribution of cluster sizes that are observed when there is no relationship between deficit scores and lesion location. Clusters from the original (true) VLSM analysis that are larger than 95\% of the null distribution of cluster sizes are taken to reflect true lesion-symptom associations (for examples of application to VLSM see \citealp{Pillay2014, Mirman2015a}). This strategy involves two separate thresholds: the first is a pre-set voxel-level $p$-threshold; the second is a permutation-based cluster size threshold. This strategy has two intuitively appealing properties. First, strokes and other neurological disorders tend to produce spatially contiguous lesions, resulting in high spatial correlations between the lesion status of neighboring voxels. Using permutation to determine a null distribution of cluster sizes intuitively controls for this spatial correlation and produces a minimum cluster size threshold that should not be observed by chance. Second, it is typical for interpretation of VLSM (and other neuroimaging) results to focus on clusters, so correcting at the cluster level (rather than the voxel level) makes this statistical strategy more closely aligned with the interpretion strategy. In the following analyses we examine this permutation-based cluster-size correction strategy for detecting true lesion-symptom relations.

\subsection{Data}

\begin{figure}[ht]
\centering
\includegraphics[width=1.0\textwidth]{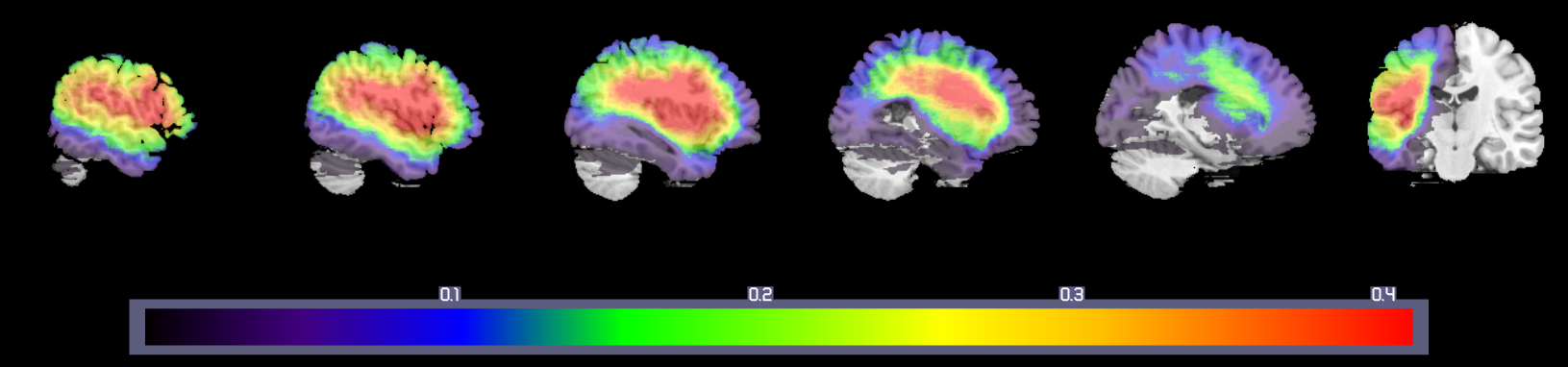}
\caption{Lesion overlap map for 124 left hemisphere stroke cases included in the present analyses. Hotter colors indicate that a larger proportion of the participant sample had lesions in that area.}
\label{fig:LesionOverlap}
\end{figure}

The lesion maps were from 124 participants with aphasia following left hemisphere stroke confirmed by computed tomography (CT) or magnetic resonance imaging (MRI) and collected as part of a larger, ongoing project investigating the anatomical basis of psycholinguistic deficits in post-acute aphasia\footnote{That project was funded by National Institutes of Health grant R01DC000191 to Myrna F. Schwartz and we are grateful to Dr. Schwartz and her team for sharing these data with us to make these analyses possible.}. The structural data were based on 108 research scans (65 MRI and 43 CT) and 16 clinical scans (5 MRI and 11 CT). Lesions imaged with MRI were manually segmented on the structural image by a trained technician and reviewed by an experienced neurologist, then registered first to a custom template constructed from images acquired on the same scanner, and then from this intermediate template to the Montreal Neurological Institute space  “Colin27” volume. Lesions imaged with CT were drawn by the experienced neurologist directly onto the Colin27 volume, after rotating (pitch only) the template to approximate the slice plane of the patient's scan. Figure \ref{fig:LesionOverlap} shows the lesion overlap map for these 124 lesion maps, which have been used in VLSM analyses reported elsewhere \citep{Mirman2015a, Mirman2015b}.

In order to have a deficit score with a known neural correlate, we calculated the percent damage in two brain regions that are widely-studied and frequently damaged in middle cerebral artery stroke aphasia: BA 45 and BA 39. An effective statistical correction strategy should approximately identify these areas (i.e., damage in BA 45 should be the  “neural correlate” of percent damage in BA 45) and the permuted data will provide additional insight into the method’s ability to reject false positives. 

\subsection{Analysis Strategy}

For each of the deficit (percent damage) scores, we conducted a basic VLSM, applied a pre-set threshold, then calculated the sizes of supra-threshold voxel clusters. We then repeated this analysis 1000 times, permuting the deficit scores for each repetition to create a random association between the scores and lesion profiles. The cluster sizes from the permutations were used to set a 95\% threshold (i.e., larger than 95\% of permutation-based clusters) for the original VLSM data. These analyses were implemented using SPM8 in Matlab and cluster sizes were computed using the {\sc bwconncomp} function from the Image Processing Toolbox. 

Six different pre-set thresholds were tested within the same set of 1000 permutations: 0.05, 0.01, 0.005, 0.001, 0.0005, 0.0001. This covers the range from the most permissive threshold (0.05) to a reasonably conservative threshold (0.0001) for initially identifying voxels for subsequent cluster size correction. The more permissive thresholds will allow more voxels into the cluster size calculation, which should produce larger clusters. Therefore, there should be a positive correlation between the pre-set $p$-threshold and the permutation-based cluster size threshold. This positive correlation is an inverse strictness relationship: more permissive $p$-thresholds produce more conservative cluster size thresholds. One motivation for this study was to examine how one might balance these inversely related factors for optimal VLSM interpretation and inference.

\subsection{Results}

\subsubsection{All Clusters}

\begin{figure}[ht]
\centering
\includegraphics[]{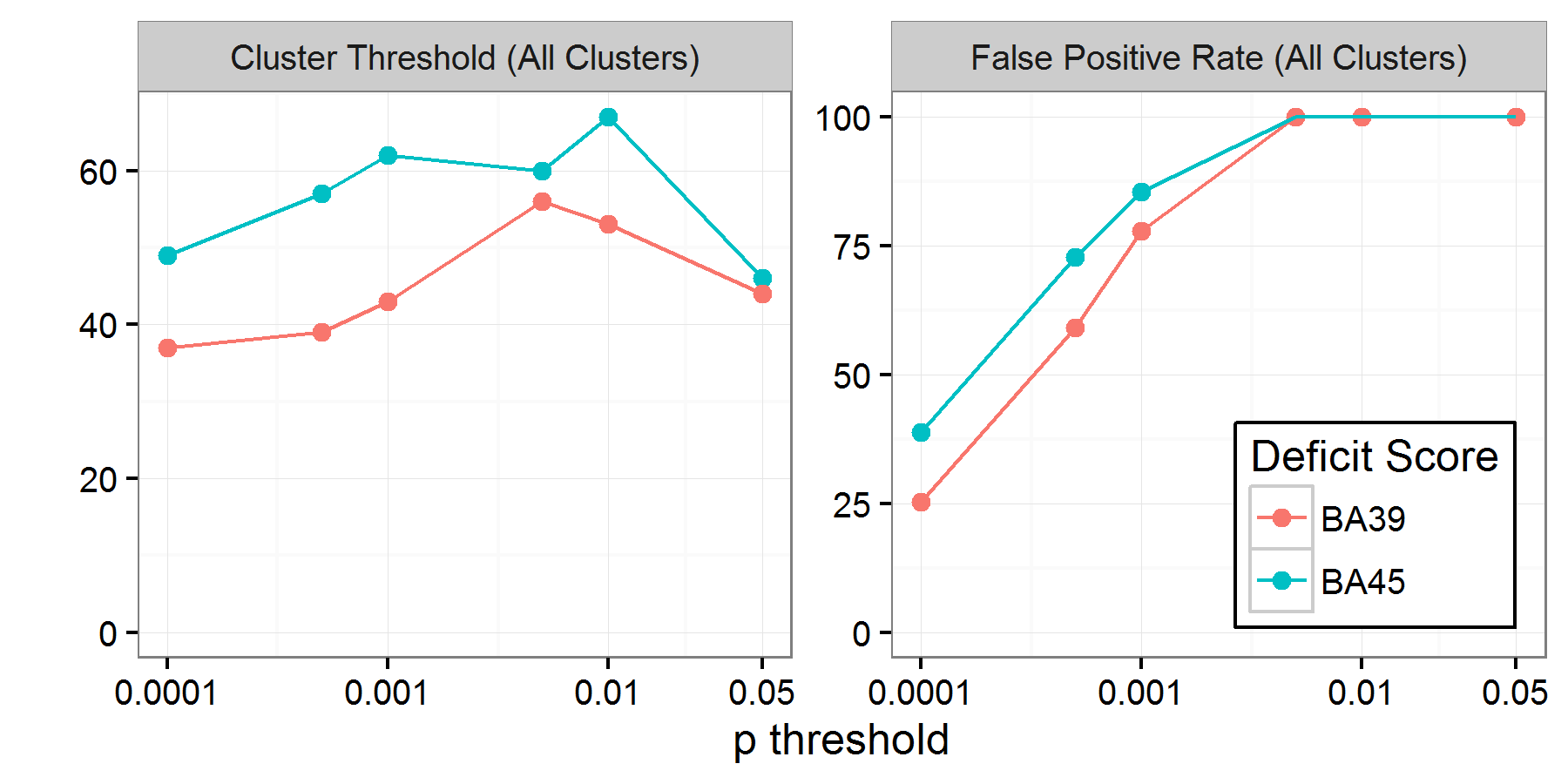}
\caption{Left panel: Cluster size thresholds based on all clusters at each $p$-threshold. Right panel: Percent of permutations with clusters larger than the cluster size threshold in the left panel (i.e., the false positive rate) at each $p$-threshold.}
\label{fig:allclust}
\end{figure}

Based on our interpretation of prior work \citep{Pillay2014, Mirman2015a}, in our first analysis, all clusters generated by each permutation were entered into the null distribution of cluster sizes for computation of the 95\% threshold. The relationship between $p$-threshold and cluster size threshold is shown in the left panel of Figure \ref{fig:allclust}. Surprisingly, there was no positive relationship between $p$-threshold and cluster size threshold – the critical cluster sizes were approximately the same across all $p$-thresholds. To examine this further, we computed the proportion of permutations that contained at least one cluster larger than the cluster size threshold. This is a measure of false positive rate in the sense that the permutations, by definition, have no systematic relationship between deficit score and lesion location. Therefore, any clusters that survive this correction would be false positives. This false positive rate is shown in the right panel of Figure \ref{fig:allclust} and was alarmingly high relative to the nominal rate of 5\%: it was 25\% for the most conservative $p$-threshold and rose to 100\% for $p \geq 0.005$.

This very high rate of false positives is particularly problematic because the cluster size threshold eliminates small clusters, so the false positives that survive this correction will be large and likely to induce unsuspecting researchers to make strong claims about lesion-deficit relationships. Using all clusters appears to match the method described by \citet{Pillay2014} and a secondary analysis reported by \citet{Mirman2015a}, but differs from the permutation-based methods for neuroimaging data analysis described by \citet{Nichols2002}, who specified that only the maximum cluster size should be used from each permutation when building the null cluster size distribution. Using only the maximum cluster size from each permutation is also analogous to using the maximum $t$-value for voxel-wise permutation-based correction. Consequently, we re-analyzed the data from the 1000 permutations reported here using only the maximum cluster size.

\subsubsection{Maximal clusters}

\begin{figure}[hb]
\centering
\includegraphics[]{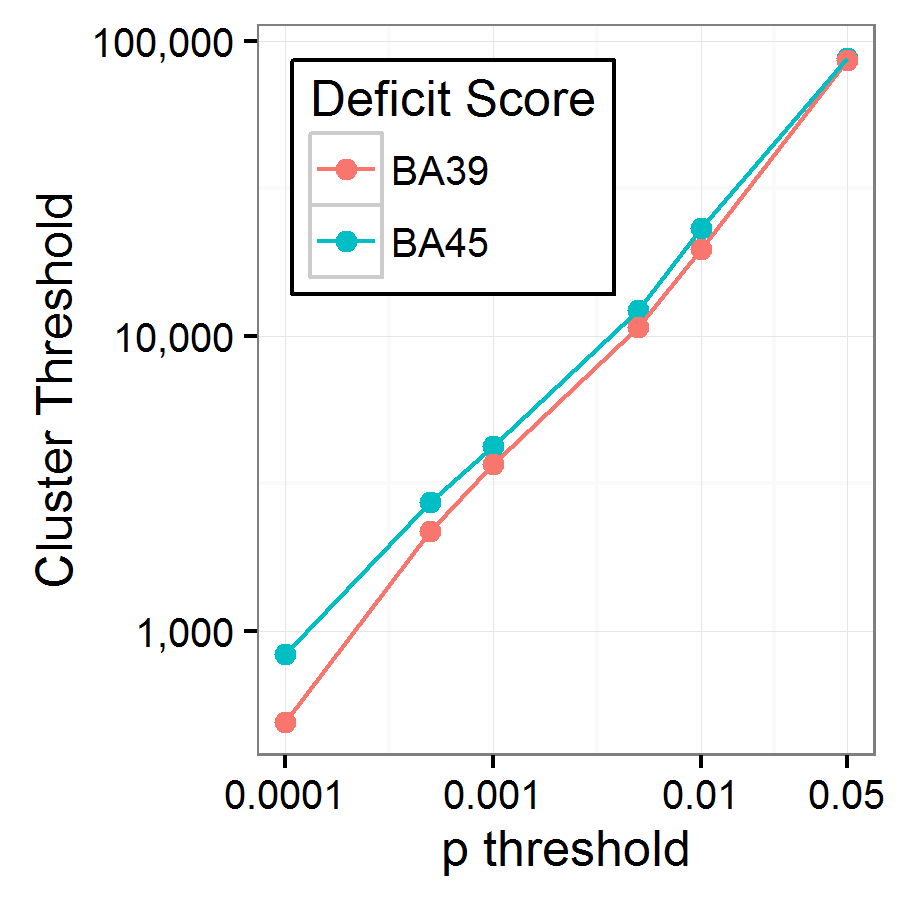}
\caption{Cluster size thresholds based on largest cluster from each permutation at each $p$-threshold. Note that both axes are logarithmically scaled.}
\label{fig:maxclust}
\end{figure}

The relationship between cluster size threshold (95th percentile of maximum cluster sizes across 1000 permutations) and $p$-threshold is shown in Figure \ref{fig:maxclust}. This analysis produced the expected positive relationship between $p$-threshold and cluster size threshold: more permissive $p$-thresholds allow more voxels into the cluster analysis, thus producing larger clusters. Indeed, the relationship is almost perfectly linear in the log-log plot in Figure \ref{fig:maxclust}. Analysis of false positives is trivial here because, by definition, 95\% of permutations had 0 clusters larger than the cluster size threshold.

The next stage was evaluating how well this method recovers the true neural correlates for each deficit score. It was immediately apparent that only the most conservative p-threshold (0.0001) produced a viable cluster size threshold under the conditions tested here. The next most conservative threshold (0.0005) produced cluster size thresholds of more than 2000 voxels; at $p$-threshold 0.01, the cluster size threshold was above 10,000 voxels. Any clusters of that size or larger would not be neuroanatomically specific enough to provide useful insights into lesion-symptom relationships.

\begin{figure}[ht]
\centering
\includegraphics[width=1.0\textwidth]{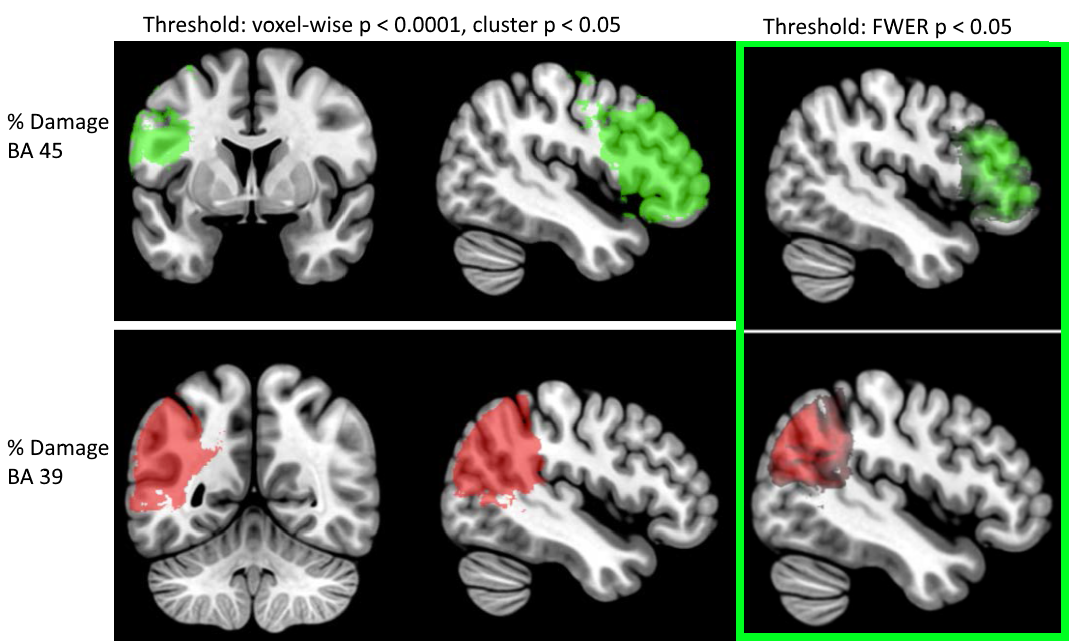}
\caption{Results of VLSM analysis of percent damage to BA 45 (top row) and BA 39 (bottom row). Left and middle columns show results thresholded at voxel-wise $p < 0.0001$ and permutation-based cluster size corrected ($p < 0.05$). Right column (outlines in green) shows results thresholded using voxel-wise permutation-based FWER at $p < 0.05$.}
\label{fig:BA39BA45}
\end{figure}

Figure \ref{fig:BA39BA45} shows the results of permutation-based cluster size correction (at voxel-wise $p < 0.0001$ and family-wise cluster size $p < 0.05$) for simulated deficit scores of percent damage in BA 45 (top row) and BA 39 (bottom row). The identified region expands beyond the bounds of the true region, covering an area that is approximately twice the size of the Brodmann Area where percent damage was used as the behavioral score. For comparison, we used the maximal $t$-value from the same 1000 permutations to compute permutation-based FWER-corrected $p < 0.05$ thresholds for each of these analyses (right column of Figure \ref{fig:BA39BA45}). This approach did a reasonably good job of identifying the critical regions.

\subsection{Discussion}

We explored the use of a permutation-based approach to determine a minumum cluster size threshold for statistical correction of VLSM. This approach is adapted from analysis of functional neuroimaging data \citep{Nichols2002} and has been previously used in VLSM \citep{Pillay2014, Mirman2015a}. Using structural lesion data from 124 participants with left hemisphere stroke, we constructed deficit scores using percent damage in BA 45 and BA 39. These behavioral scores were used to assess the permutation-based cluster size correction method on three criteria (1) correctly detecting the relationship between damage to BA 45 or BA 39 and the corresponding deficit score, (2) falsely detecting clusters in the permutations (which, by definition, have no lesion-deficit relationships), and (3) incorrectly detecting voxels outside the critical BA 45 and BA 39 regions. 

Our first discovery was that it is absolutely critical that only the largest cluster from each permutation is included in the null distribution of cluster sizes -- using all clusters produces an extraordinarily high rate of false positive clusters in the permuted data. In fact, this was a \emph{re}-discovery because \citet{Nichols2002} specified that only the largest cluster should be used, though it is not clear whether this was the case in prior applications of this method to VLSM \citep{Pillay2014}. 

When only the largest cluster was used, the false positive rate was controlled (i.e., 95\% of all permutations had no clusters that passed the cluster size threshold) and the critical BA regions were correctly identified. However, the supra-threshold clusters extended well beyond the boundaries of the correct BA regions. This pattern suggests that permutation-based cluster size approach can correctly reject cases in which there is no consistent relationship between behavioral score and lesion location. However, it appears to be insufficiently spatially specific when a true relationship exists -- if a lesion-symptom relationship does exist, this method will detect the correct region but spatially contiguous regions will also be included in the critical cluster. This spill-over effect may lead to incorrect interpretation of the results, so a better method is needed. As shown in the right column of Figure \ref{fig:BA39BA45}, permutation-based FWER was more effective at identifying the correct (simulated) lesion-symptom relationship. However, percent damage is a very strong relationship and, as discussed in the Introduction, this FWER method is very conservative because it controls the possibility of a single false-positive voxel. This single-voxel standard does not align with how VLSM results are interpreted and this conservatism carries real costs for scientific progress. In the next section we explore a generalization of this approach that allows balancing false positives against false negatives and transparently reporting the evidence.

\section{Continuous Permutation-Based FWER}

The standard permutation-based FWER correction method proceeds as follows: (1) permute behavioral data and conduct VLSM analysis, (2) identify the maximal test statistic (typically, the most extreme $t$-value), (3) repeat steps 1 and 2 many times to build a null distribution of maximal $t$-values, (4) compute the $n$-th percentile of that null distribution to determine a threshold for the test statistic, which corresponds to $n$\% of the permutations having 0 voxels that exceed this threshold. A typical value of $n$ is 95, which produces a FWER-corrected $p < 0.05$: less than 5\% of the permutations had even a single voxel that exceeded this $t$-threshold.

\begin{figure}[ht]
\centering
\includegraphics[width=1.0\textwidth]{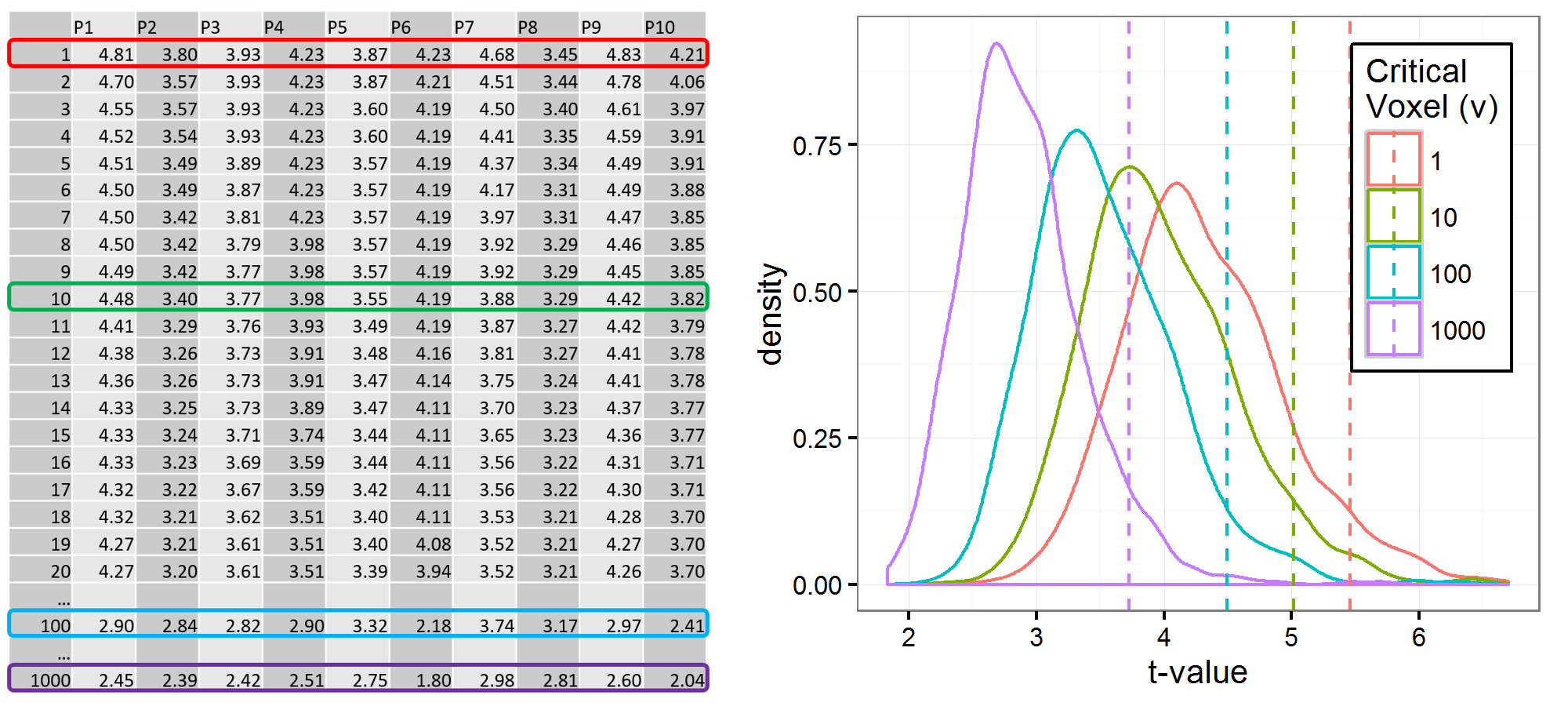}
\caption{Example of continuous FWER threshold calculation. Left panel shows the top sorted $t$-values from the first 10 permutations. The red box highlights the standard $v=1$ permutation $t$-values, which produce the red distribution in the right panel and the 95\% threshold indicated by the red dashed line. The green box highlights the $v=10$ permutation $t$-values, which produce the green distribution in the right panel and the 95\% threshold indicated by the green dashed line. Analogous $t$-values, distributions, and 95\% thresholds are also shown for $v=100$ (blue) and $v=1000$ (purple).}
\label{fig:cFWER_ex}
\end{figure}

This approach controls the rate of single-voxel false positives, but it can be generalized to multi-voxel false positives by focusing on the $v$-th most extreme test statistic. The standard strategy is the special case when $v=1$, thus using the most extreme voxel-wise test statistic from each permutation, and controlling the rate of 1 false positive voxel. If, for example, $v=10$, one would similarly use the 10th most extreme voxel-wise test statistic from each permutation, and control the rate of up to 10 false positive voxels. An example is shown in Figure \ref{fig:cFWER_ex} where the left panel shows the sorted $t$-values from the first 10 permutations and the right panel shows permutation-based $t$-value distributions and 95\% thresholds at $v=1$ (red), $v=10$ (green), $v=100$ (blue), and $v=1000$ (purple). A single set of permutations produced the set of possible 95\% thresholds at different critical voxel ($v$) values, reflecting the expected number of false positive voxels at the corresponding $t$ threshold. Not surprisingly, that $t$ threshold decreases as the $v$ value increases, but the relationship remains the same as the standard FWER case: the $v$ value specifies the maximum number of voxels that exceeded the corresponding $t$-threshold in 95\% of the permutations. That is, a reasonable upper bound on number of false positive voxels at that $t$ threshold. 

Since interpretation of VLSM results typically relies on a large set of voxels, 10 false positive voxels are unlikely to affect interpretation. This is not to say that $v=10$ should be adopted instead of $v=1$; rather, this generalization would allow investigators to assess the strength of their evidence in a flexible way and to report that assessment in a way that would allow readers (and reviewers) to evaluate the claims. We refer to this extension of the standard permutation-based FWER as \emph{continuous} permutation-based FWER because it uses the same permutation-based FWER strategy but allows values of $v>1$.

Allowing multiple false positive voxels and quantifying that rate of false positives makes continuous permutation-based FWER somewhat similar to the false discovery rate (FDR) approach, but with two important differences. First, FDR is designed to control the \emph{proportion} of supra-threshold voxels that are expected to be false positives (this proportion is usually reported as the \emph{q}-value), whereas the FWER approach quantifies the \emph{number} of possible false positive voxels, which may be a high or low proportion of supra-threshold voxels. Second, continuous FWER is permutation-based, which means that (unlike FDR) it makes no assumptions about distributions of data or test statistics. The latter property is important because lesion data may violate the assumptions of FDR severely enough to make FDR unreliable for VLSM. Here we report results from application of this continuous permutation-based FWER approach in several contexts. In addition, we used the permutation data to evaluate whether the nominal FDR $q$-value correctly quantifies the proportion of supra-threshold voxels that are expected to be false positives.

\subsection{Analysis Strategy}

We conducted analogous analyses on three sets of data. The first was the same data as in the cluster size threshold analyses above: 124 left hemisphere stroke cases with two simulated behavioral scores, percent damage in BA 45 and BA 39. This provides a relatively large data set with a known correct outcome. The second was randomly sampled sub-sets of these data to examine how continuous FWER and FDR perform for smaller data sets. We used 50 random half-samples (N=62) and 50 random quarter-samples (N=31). The third data set was speech recognition deficit data from 99 left hemisphere stroke cases reported in a recent article \citep{Mirman2015a}. This data set provided an opportunity to test continuous FWER and FDR in the context of real behavioral data where the outcome was relatively uncontroversial: deficits in speech perception and spoken word recognition should be associated with lesions in left superior temporal lobe regions. Although not quite as certain as using simulated behavioral scores, this outcome is very strongly expected and using real behavioral data allowed us to test these statistical methods in the context real-world variability. 

For each analysis, we conducted standard VLSM analysis and computed continuous permutation-based FWER 95th percentile thresholds at $v = 1, 10, 100, 1000$ based on 1000 permutations. That is, $t$-value thresholds where 95\% of the permutations had fewer 1, 10, 100, or 1000 supra-threshold voxels. We then computed the number of voxels in the original VLSM that had $t$ values greater than the $t$ threshold at each $v$ threshold, which is the number of FWER-corrected ($p<0.05$) voxels at each $v$ threshold. The $v$ threshold and the number of supra-threshold voxels were then used to compute an \emph{effective} $q$ value: the proportion of supra-threshold voxels that can be expected to be false positives based on the $v$ value. For example, if 500 voxels survived the correction at $v=10$, that would correspond to $q = 10/500 = 0.02$. This effective $q$ value was then used to compute a FDR-corrected $t$-threshold in order to test it against the corresponding permutation-based $t$-threshold and evaluate whether the nominal proportion of false positives ($q$) was correct. These analyses were carried out in R version 3.2.4 \citep{R} using the ANTsR package version 0.3.3 \citep{ANTsR} and the FDR implementation in the AnalyzeFMRI package version 1.1-16 \citep{AnalyzeFMRI}.

\subsection{Results}

\subsubsection{Simulated scores, full sample}

\begin{figure}[ht]
\centering
\includegraphics[width=1.0\textwidth]{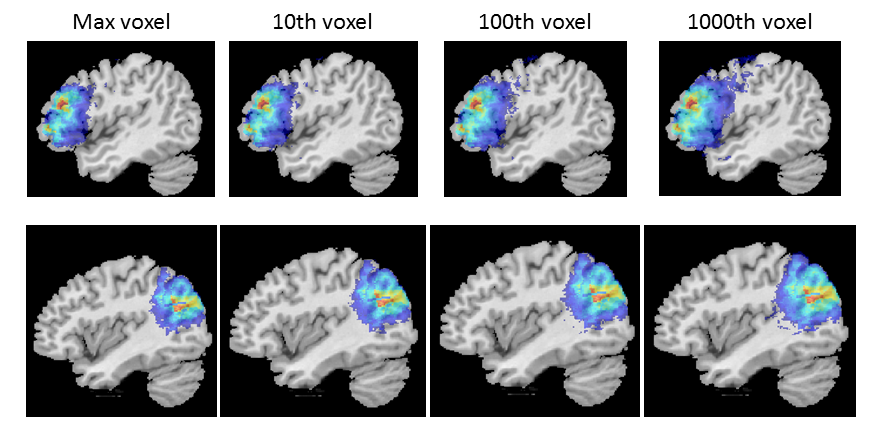}
\caption{Results of VLSM analysis of percent damage to BA 45 (top row, at $x=45$) and BA 39 (bottom row, at $x=50$). Thresholded using permutation-based continuous FWER at $p < 0.05$.}
\label{fig:continFWER_BA}
\end{figure}

Figure \ref{fig:continFWER_BA} shows the VLSM results corrected at $p<0.05$ using continuous permutation-based FWER with $v=1,10,100,1000$ voxels. The first column in Figure \ref{fig:continFWER_BA} corresponds to the standard FWER correction, which is also shown in the right column of Figure \ref{fig:BA39BA45}. The other columns show that (unsurprisingly) the supra-threshold region increases as the number of allowed false positive voxels increases.

\begin{figure}[ht]
\centering
\includegraphics[]{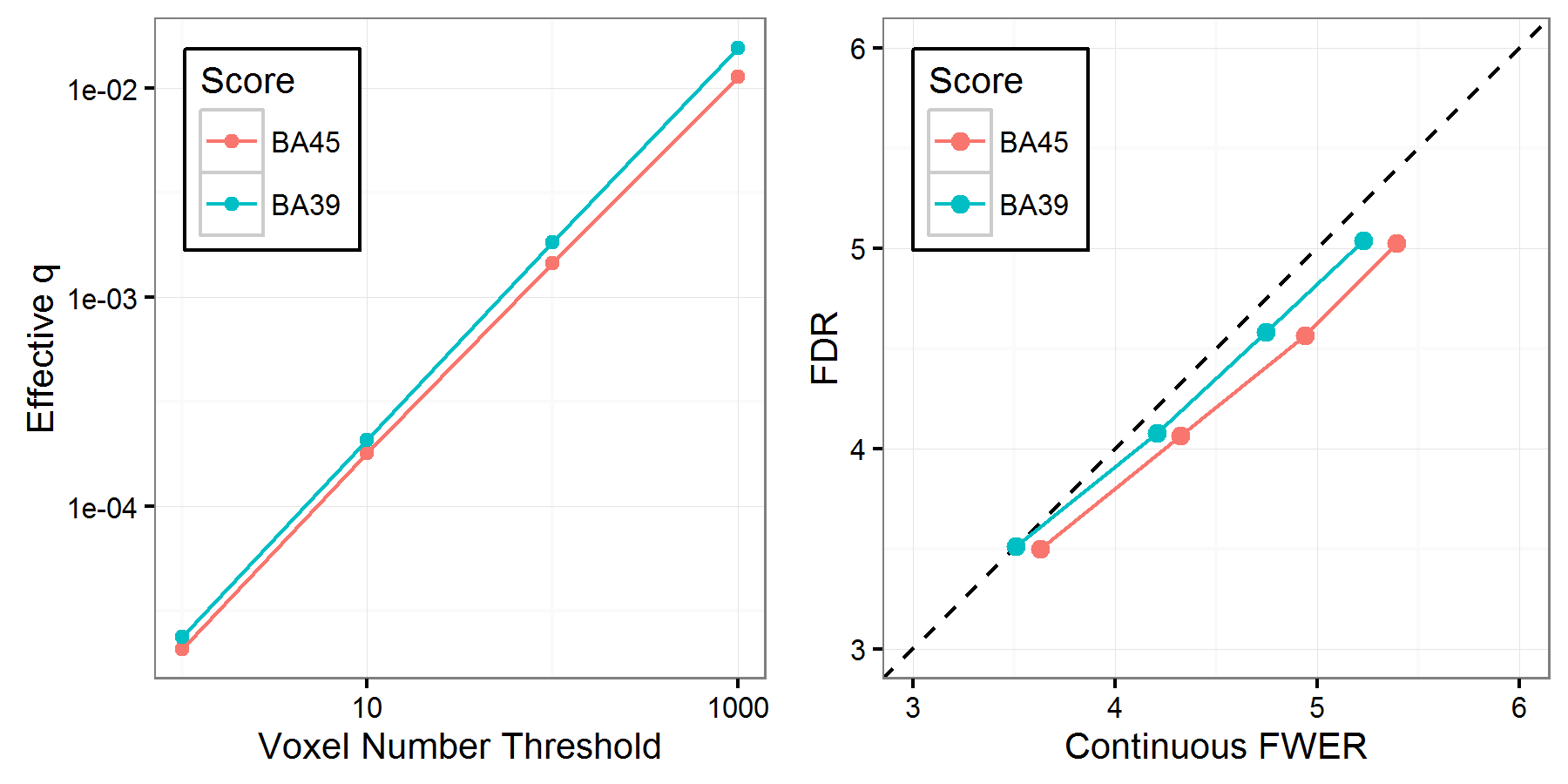}
\caption{Left: Relationship between voxel number threshold ($v$) and the proportion of supra-threshold voxels that can be expected to be false positives (effective $q$). Right: Relationship between critical $t$-values ($t$-thresholds) determined by continuous FWER correction and FDR correction.}
\label{fig:continFWER_BA_lines}
\end{figure}

The left panel of Figure \ref{fig:continFWER_BA_lines} shows the relationship between the voxel number threshold ($v$) in the continuous FWER correction and the resulting effective $q$ value. The points in the bottom left corner correspond to the standard, $v=1$, FWER correction. As $v$ was increased, there was a corresponding increase in $q$, the proportion of supra-threshold voxels that can be expected to be false positives. This relationshp between $v$ and effective $q$ was essentially linear (on log-log scale) and virtually identical for the BA45 and BA39 test cases. Note that even at the most lenient threshold, $v=1000$, the effective $q$ value was still quite low (0.011 for BA45; 0.016 for BA39), presumably because of the very strong relationship between percent damage in a BA (the simulated deficit score) and lesion in that region.

The right panel of Figure \ref{fig:continFWER_BA_lines} shows the relationship between the critical $t$-value (i.e., the corrected $t$-threshold) as computed by continuous FWER and by FDR. The FDR-corrected $t$-value was computed using the effective $q$ from the left panel of Figure \ref{fig:continFWER_BA_lines}. Since effective $q$ is a non-parametric estimate of the true rate of false positive voxels at the corresponding $t$-threshold, if FDR works as intended, then it should produce $t$-thresholds that are very similar to those computed by the non-parametric continuous FWER method. This is the pattern shown in the right panel of Figure \ref{fig:continFWER_BA_lines}: virtually identical critical $t$-thresholds computed by FDR correction and by continuous FWER correction, for both BA45 and BA39 scores. That is, the permutations confirm that, at $q=0.01$, FDR correction accurately produced a critical $t$-value such that up to 1\% of supra-threshold voxels could be expected to be false positives. 

This is an encouraging result for application of FDR to VLSM data because it shows the $q$-value correctly quantifies the proportion of false positive voxels. However, this is an ideal scenario in at least two ways: (1) a very strong relationship between simulated score (percent damage in BA 45 or BA 39) and lesion location, and (2) a relatively large sample size (N=124). To evaluate the contribution of this second factor we conducted further analyses of these same data but using smaller sub-samples of the data.

\subsubsection{Simulated scores, sub-samples}

\begin{figure}[ht]
\centering
\includegraphics[]{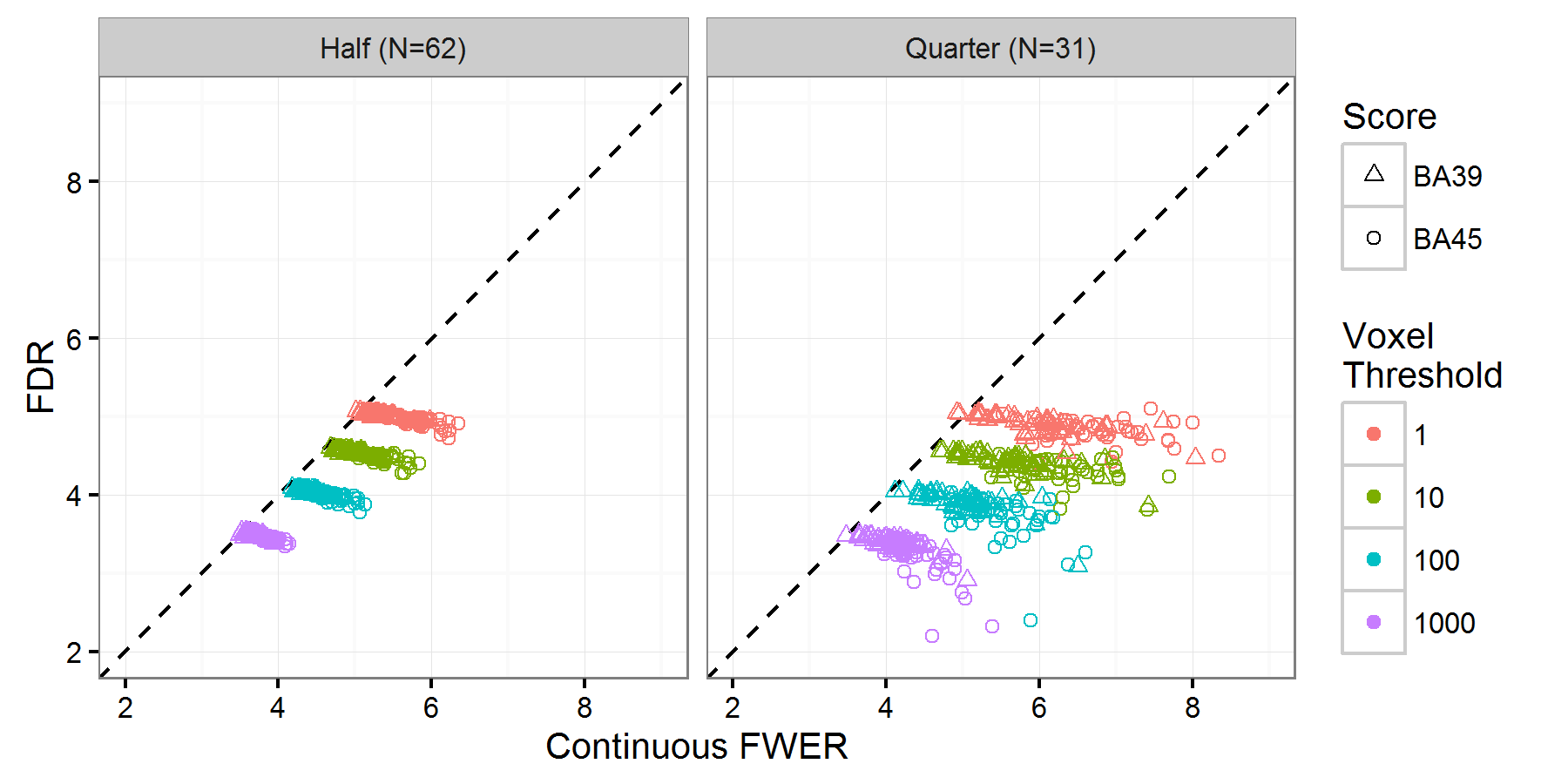}
\caption{Relationship between critical $t$-values ($t$-thresholds) determined by continuous FWER correction and FDR correction for 50 randomly selected half-samples (left panel) and 50 quarter-samples (right panel). Deficit scores are percent damage in BA39 (triangles) or BA45 (circles). Each point represents one of the random sub-samples.}
\label{fig:subSamples}
\end{figure}

The initial analysis of 124 participants constitutes a fairly large sample size by VLSM standards. More modest sample sizes (e.g., 40-60) are far more common and many studies report even smaller samples (e.g., 20-40). Smaller sample sizes are more likely to (more severely) violate assumptions of FDR, so, although FDR worked as intended for the full N=124 sample, it may not be robust at smaller sample sizes. In particular, the spatial coherence of lesions means that the voxel-wise tests violate the test independence assumption and symptom scores are often non-normal -- both of these problems will tend to be more severe for smaller sample sizes. However, FDR is robust to some degree of assumption violation \citep{Groppe2011a, Groppe2011b}, so it may produce approximately correct results even under these conditions. To evaluate this, we repeated the comparison of continuous FWER and FDR using 50 half (N=62) and 50 quarter (N=31) random sub-samples of the full data set. Figure \ref{fig:subSamples} shows scatterplots of the critical $t$-values based on continuous FWER (at each of the four $v$ thresholds) and the corresponding FDR critical $t$-value computed using the effective $q$ value. The dashed line represents exact equivalence between continuous FWER and FDR, which was approximately true for the full sample. For these smaller samples, the FDR method tends to produce a substantially less conservative critical $t$-value threshold for each effective $q$ value. There was substantial variability in how far the FDR threshold deviated from the continuous FWER threshold, with some sub-samples showing fairly close correspondence (as was observed for the full sample), but most falling short of that. Comparing the half-sample and quarter-sample data (left vs. right panels in Figure \ref{fig:subSamples}) shows that both the departure of FDR from continuous FWER and the variability of their relation became more extreme for smaller samples. 

The interpretation of continuous FWER is transparent, so this discrepancy between the methods represents a problem for FDR. For example, for the first quarter-sample in the BA 45 analysis, the (traditional) $v=1$ FWER threshold produced a critical $t$-value of 6.0 and 5904 voxels had $t$-values above that threshold. That is, permutation analysis indicates that only 1 out of those 5904 can be expected to be a false positive, which is an effective $q = 1/5904 = 0.00017$. Applying FDR to these data with $q=0.00017$ produced a critical $t$-value of 4.9 and 11,429 voxels passed that $t$-threshold. According to FDR, only 0.017\% of those 11,429 voxels are expected to be false positives, which is approximately 2 voxels. However, the permutation data reveal that, if there were no relationship (i.e., if the null hypothesis were true), then approximately 100 voxels could be expected to exceed a critical $t$-value of 4.9; about 50 times more than the nominal rate implied by the $q$-value.

\begin{figure}[ht]
\centering
\includegraphics[]{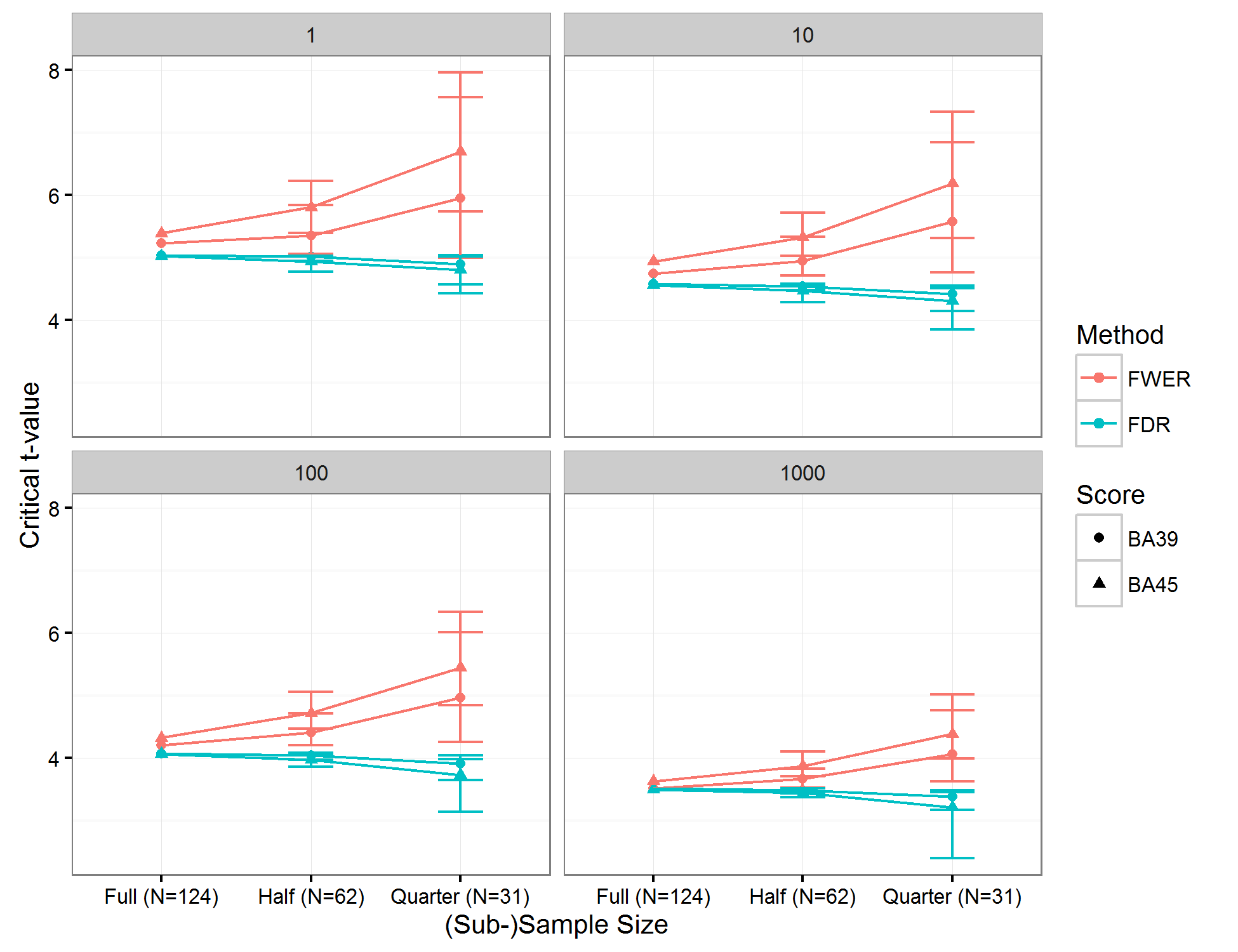}
\caption{Critical $t$-values ($t$-thresholds) determined by continuous FWER correction and FDR correction as a function of sample size. Deficit scores are percent damage in BA39 (triangles) or BA45 (circles). Error bars represent 95\% confidence intervals.}
\label{fig:sampleSize}
\end{figure}

Sample size appears to have different effects on FWER correction and FDR correction (see Figure \ref{fig:sampleSize}). FDR-corrected thresholds are relatively constant across sample sizes, but FWER-corrected thresholds increase as sample size becomes smaller. This pattern may arise because violations of assumptions have a bigger effect in smaller samples. Permutation-based approaches avoid making assumptions and directly compute $p$-values while maintaining distributional properties of the original data through all of the permutations. In contrast, the FDR computation is based on just the distribution of \emph{voxel-level test statistics}, without any information about sample size and with an assumed distribution of the data. As a result, FWER becomes more conservative at smaller sample sizes, but FDR does not. This observation is particularly important for the many VLSM studies that have sample sizes in the 30-60 range: FDR may produce anti-conservative results for these smaller sample sizes, but the conservative standard ($v=1$) FWER correction may relegate these studies to the file drawer. We return to this issue further after examining a real deficit example. 

\subsubsection{Speech recognition scores}

All of the preceding analyses used simulated deficit scores that had rather strong lesion-symptom relations. A strong signal is easy to detect and may obscure weaknesses of a statistical method, so it is important to test statistical methods with more realistic data. Adding noise to simulated lesion-symptom relations would effectively weaken them, but real lesion-symptom relations are not simply randomly noisy, so there is no guarantee that adding random noise would capture the ways that real lesion-symptom relations differ from simulated ones. However, using real deficit data is somewhat risky because the true lesion-symptom relation is not known. To mitigate this concern, we chose a relatively uncontroversial case: composite speech recognition deficit scores determined by a factor analysis of data from 99 individuals with left hemisphere stroke \citep{Mirman2015a}. These scores primarily reflect phoneme discrimination and auditory lexical decision performance (for details see  \citealp{Mirman2015a, Mirman2015b}) and it is quite well-established that these tasks primarily engage left superior temporal lobe structures (e.g., \citealp{Hickok2015, DeWitt2012}). As a result, this data set allows us to investigate how continuous FWER and FDR would work in a real VLSM context while being fairly confident about what the correct result should be.

\begin{figure}[ht]
\centering
\includegraphics[width=1.0\textwidth]{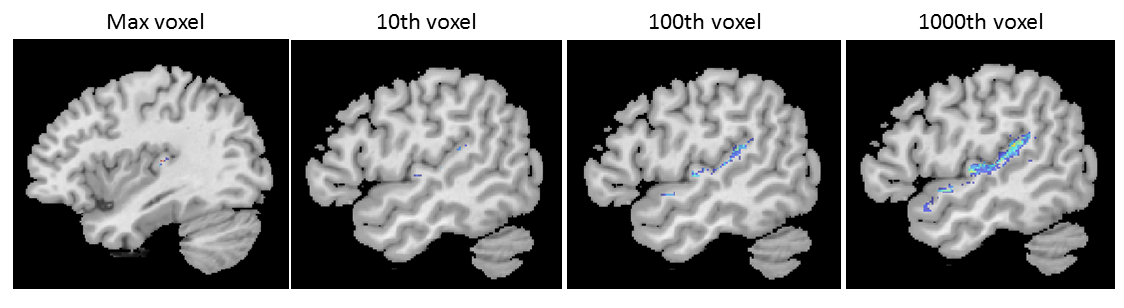}
\caption{Results of VLSM analysis of Speech Recognition scores. Thresholded using permutation-based continuous FWER at $p < 0.05$, panels show results at different $v$ thresholds: $v=1, 10, 100, 1000$. Note: at $v=1$, the small set of supra-threshold voxels was located more medially than the main group of voxels in the other panels, so the left panel shows the results at $x=53$ while the other panels are at $x=40$.}
\label{fig:SpeechRec}
\end{figure}

Figure \ref{fig:SpeechRec} shows the VLSM results after continuous FWER correction at $v=1, 10, 100, 1000$. As expected, the identified region is in the superior temporal lobe and, as in the simulated scores analyses, a more relaxed $v$ threshold produces a larger supra-threshold region. At the standard, $v=1$, threshold, the FWER corrected critical $t$-value was 5.45 and 57 voxels passed this threshold. On one hand, this is a positive result: it is unlikely ($p<0.05$) that even one of those 57 voxels is a false positive, so we should feel confident about interpreting those 57 voxels as being critically important for speech recognition. On the other hand, those 57 voxels are virtually invisible in the figure (left-most panel in Figure \ref{fig:SpeechRec}; even the 261 voxels that passed the $v=10$ threshold of $t=5.04$ are hard to see) and it seems unlikely that editors, reviewers, and readers would be convinced by a 57-voxel result (or even a 261-voxel result). Such a small cluster might even be within the margin of error of the lesion segmentation algorithms and the warping algorithm used to align individual lesion maps to a common template for analysis. Relaxing the $v$ threshold reveals an easier to interpret result. For example, at $v=100$, the critical $t$-value was 4.42 and 1527 voxels passed this threshold. Up to 100 of those 1527 voxels (6.5\%) can be reasonably expected to be false positives, but that probably would not affect how one would interpret the result in the $v=100$ panel of Figure \ref{fig:SpeechRec}. 

This is not to say that the threshold should be moved from $v=1$ to $v=100$ -- there may circumstances where 100 voxels (or 6.5\% of the supra-threshold voxels) \emph{would} affect the interpretation of VLSM results. A flexible $v$ threshold gives the continuous FWER approach two important advantages. First, researchers can select the $v$ threshold that is most appropriate for testing their hypothesis and can report their results at multiple $v$ thresholds. If the evidence is strong, they can draw strong conclusions; if the evidence is not so strong, they can draw tentative conclusions. For example, the small anterior-most cluster of voxels that passed the $v=100$ threshold may be smaller than 100 voxels and disappears at the $v=10$ threshold. This is relatively weak evidence that anterior superior temporal regions are critical for speech recognition, especially compared to the much stronger evidence that posterior superior temporal regions are critical for speech recognition. This is importantly different from the standard (and increasingly criticized) dichotomous logic that an effect is either “significant” or non-existent (see also \citealp{Chen2016}). Second, the likely upper limit of false positive voxels is transparently available to the reviewers and readers, who can then evaluate how the $v$ threshold influences the conclusions; for example, whether the possibilty of 100 false positive voxels undermines the conclusions or not. Transparently reporting the strength of the evidence allows the science to accumulate -- multiple studies that weakly or tentatively show the same pattern can be aggregated to strongly support a conclusion, and contradictory results can be evaluated on the strength of their evidence.

\begin{figure}[ht]
\centering
\includegraphics[width=0.65\textwidth]{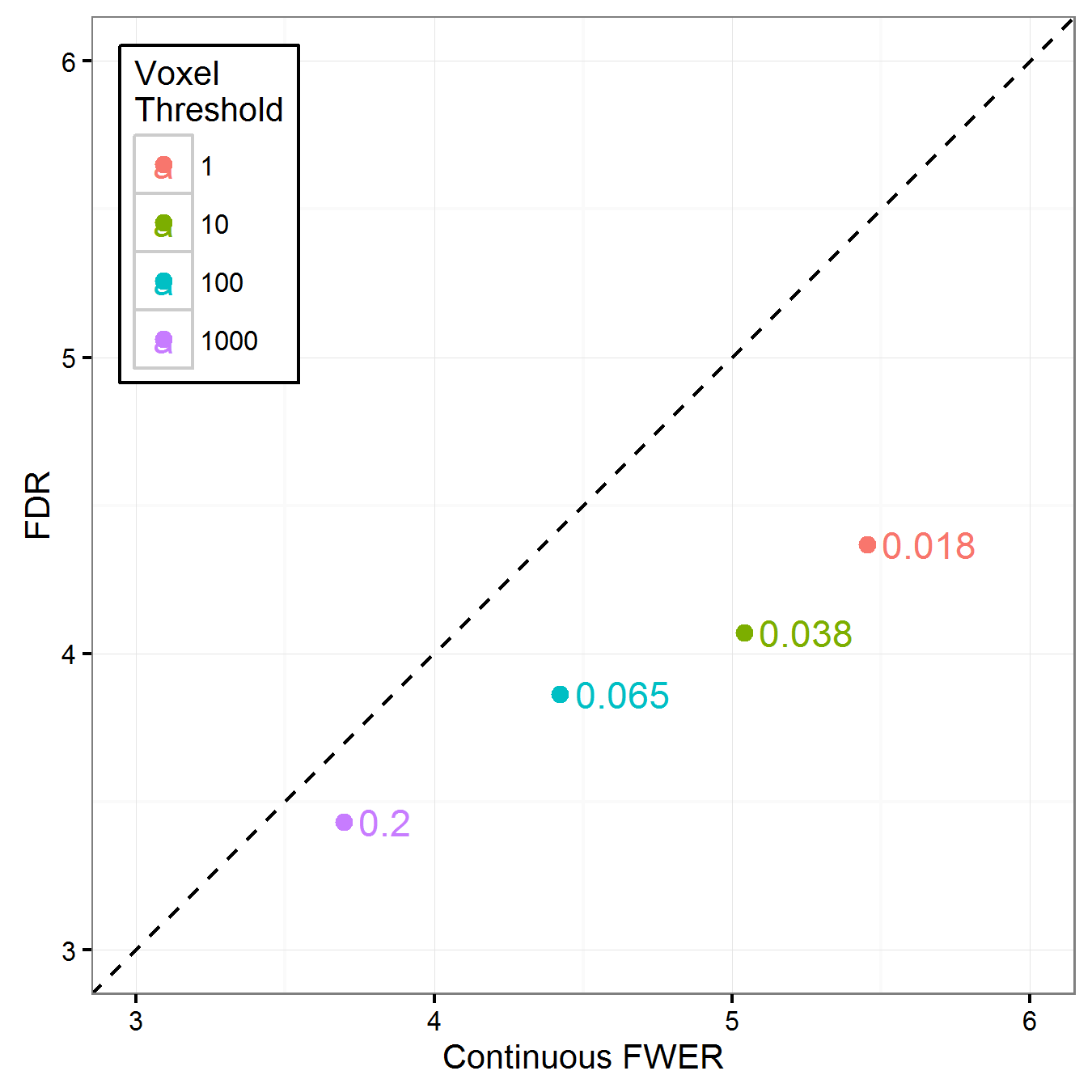}
\caption{Critical $t$-values ($t$-thresholds) determined by continuous FWER correction (at $v = 1, 10, 100, 1000$) and FDR correction for analysis of speech recognition scores. Points indicate the critical $t$-values, numbers next to the points indicate corresponding $q$-values.}
\label{fig:SpeechRec_FWER-FDR}
\end{figure}

Figure \ref{fig:SpeechRec_FWER-FDR} shows the relationship between $t$-value thresholds based on continuous FWER correction and FDR correction for the speech recognition data. As in the sub-sample analyses, continuous FWER is consistently more conservative than FDR, indicating that FDR produced incorrect results. For example, at $q=0.018$, the FDR-corrected critical $t$-value was 4.37 and 1703 voxels passed that threshold. The nominal expectation is that up to 1.8\% of those voxels may be false positives (about 30 voxels), but the permutation data indicate that more than 100 voxels can be expected to be false positives, or more than 3 times higher than expected. As in the sub-sample analyses, this result suggests that researchers should be wary of using FDR correction with VLSM data.

\subsection{Discussion}

Permutation-based FWER correction uses permutations of the observed data to build a null distribution of voxel-wise test statistics, then uses this null distribution to set thresholds for evaluating the test statistics in the original (true) analysis. The standard version of this approach uses only the most extreme voxel-wise test statistic from each permutation and the resulting threshold makes it unlikely that even a single false positive voxel will be observed. We examined a generalization of this approach in which the threshold is not based on only the most extreme ($v=1$) test statistic. Using the $v$-th most extreme test statistic (where $v > 1$) provides a way to quantify the possible rate of false positive voxels: up to $v$ voxels can be reasonably expected to be false positives. We refer to this approach as \emph{continuous permutation-based FWER correction}. Since single voxels rarely (if ever) affect interpretation of VLSM results, this extension aligns the correction method with how VLSM results are interpreted, and allows for transparently reporting the strength of the evidence. Analyses of speech recognition deficit scores provided a particularly clear demonstration of the value of quantifying rates of false positive voxels within a flexible framework. At the standard $v=1$ threshold, only 57 voxels survived the correction -- a statistically “significant” result that is hard to interpret. Examining the data at $v=10$, 100, and 1000 revealed a clear (and unsurprising) relationship between posterior superior temporal lobe damage and speech recognition deficits. Researchers can calibrate the $v$ threshold to their hypotheses or regions of interest. For example, a hypothesis about a very specific region (e.g., role of area Spt in speech processing, \citealp{Rogalsky2015}) might require a small $v$ threshold, whereas a hypothesis about a broader region (e.g., role of the inferior frontal gyrus in lexical selection, \citealp{Harvey2015, Mirman2013}) might allow a larger $v$ threshold. In addition to providing researchers with more flexibility in evaluation of their data, this approach provides an intuitive way to report the expected upper limit of false positive voxels, which allows readers to evaluate the conclusions as well.

Importantly, continuous permutation-based FWER maintains the advantages of the standard non-parameteric permutation-based FWER correction strategy. These advantages became apparent in the comparison between continuous FWER correction and FDR correction. Like continuous FWER, FDR (nominally) quantifies the expected rate of false positive voxels. For a relatively large sample (N=124) with a very strong simulated lesion-symptom relation, FDR quite accurately quantified the rate of false positive voxels. However, at smaller sample sizes (N=62, N=31) and with real deficit data (presumably less strong lesion-symptom relation), FDR consistently under-estimated the rate of false positives. To our knowledge, this is the first concrete evidence that FDR correction may not be appropriate for VLSM analysis.

\section{General Discussion and Conclusions}

Permutation-based FWER correction is the current “gold standard” correction for multiple comparisons in VLSM. The main weakness of this approach is that it aims to control the occurrence of even a single false positive voxel, which is not the scale at which VLSM results are interpreted. We described an extension - continuous permutation-based FWER - which better aligns with VLSM interpetation and allows researchers a more flexible balance between false positives and false negatives. The formulation of continuous FWER also provides a way to transparently report the upper limit of the expected number of false positive voxels. The standard permutation-based FWER strategy is to build a null distribution using only most extreme voxel-wise test statistic from each permutation. Continuous FWER uses the $v$-th most extreme voxel-wise test statistic, so the standard approach is the special case where $v=1$, but other values $v>1$ may be used as appropriate for a particular data set and hypothesis. This provides a principled way for researchers to flexibily set the upper limit of how many false positive voxels are allowed and to report this limit along with their results, so readers can also evaluate the evidence. Since single voxels rarely (if ever) affect the interpretation of VLSM results, this flexibility lets researchers align their statistical method with their interpretations of the results.

In addition to continuous FWER, we examined two other methods of correction, but those results were not encouraging. Using permutations to set a minimum cluster size tended to produce clusters that extended well beyond the correct region. This may have been partly due to a true correlation between damage in adjacent regions and damage in the correct region (i.e., spatial coherence); however, since the other correction methods seemed to avoid this spill-over problem (or at least mitigate it), this does not appear to be the primary explanation. Instead, we suspect this spill-over occurred because weak or noisy effects in adjacent voxels were incorporated into true clusters, with the unfortunate consequence of blurring the boundary of the true symptom-related region. This method appeared to be effective at controlling the occurrence of false positive clusters, but the spill-over effect poses a problem for identifying brain-behavior relationships. The false discovery rate (FDR) approach is inferentially similar to continuous FWER; it aims to quantify the rate of false positive voxels. FDR performed quite well for larger samples with strong lesion-symptom relations, but consistently underestimated the rate of false positive voxels when the sample sizes were smaller and in a real data case (where the lesion-symptom relation is likely to be weaker). This suggests that researchers should be wary of using FDR in conventional VLSM analyses.

There is an inherent trade-off between false positives and false negatives: striving to eliminate false positives will necessarily result in missing many true effects, but generalizing from every observation will necessarily produce some incorrect inferences. Setting arbitrary thresholds of statistical significance makes evidence appear more dichotmous than it really is; statistical thresholds encourage binary thinking in which an effect is either significant or non-existent. This dichotomy is further exacerbated by publication bias because weaker, not statistically significant results are simply not published. This reifies the sense that effects that do not pass the significance threshold are non-existent, leading to a biased scientific literature and undermining evidence accumulation. Balancing false positives and false negatives is particularly challenging in VLSM, where participant recruitment and testing is difficult and relatively expensive, and samples are generally large relative to typical research in neuropsychology and cognitive neuroscience. A typical VLSM research project might require a long period of expensive data collection to reach a reasonable sample size of, say, N=50. If analysis of that data set produced results just short of the standard FWER $v=1$ statistical threshold, the researchers would be left with an unpublishable result. Substantially increasing the sample size is likely to be impractical (and perhaps impossible) as it would require another long, labor-intensive, and expensive data collection effort. Addressing this challenge requires a statistical correction method that allows researchers to flexibly balance false positives and false negatives and to report how they struck that balance in a transparent fashion so that readers can interpret the evidence. The $v$ threshold plays this role in the continuous permutation-based FWER correction method: $v$ is the expected upper limit of false positive voxels, which can be adjusted to suit the researchers' hypotheses and reported for readers to use in their evaluation.

It may be possible to further improve the correction methods described here by considering unique “patches” (where lesion status is perfectly correlated across participants) rather than individual voxels \citep{Kimberg2007}. Our (unsystematic) comparisons suggest that, in a typical data set, the number of unique patches may be an order of magnitude smaller than the number of voxels. Such a vast reduction in the number of (redundant) tests would have substantial consequences for these methods, as well as for the processing time required to compute them. More generally, including voxel neighborhood information as part of voxel-level corrections may lead to even more effective algorithms. In addition, recent development of multivariate lesion-symptom mapping methods \citep{Zhang2014}, which evaluate lesion-symptom relationships across all voxels simultaneously, provide a better method for studying brain-behavior relationships and mitigate the need for multiple comparisons correction. Such methods are not in wide use yet, but they offer a promising alternative approach.

VLSM is important for basic and translational human neuroscience. Analysis of lesion-symptom relations has been at the core of cognitive neuroscience research since the mid-19th century and remains critical to the field. Lesions that produce chronic deficits in a particular domain or task provide the strongest evidence that the damaged neural structures were critical for that domain or task. This method is an important complement to functional neuroimaging in neurologically-intact populations, but the differences in data collection challenges create somewhat different statistical demands. Cognitive neuroscience has tremendous potential for stimulating development of new and improved diagnosis, treatment, rehabilitation, and education strategies. That potential cannot be realized without testing the affected populations. VLSM offers a unique opportunity for research that answers fundamental questions about the neural basis of cognition while addressing the real-world problem of understanding neurogenic cognitive deficits. Robust, flexible, and transparent statistical methods play an important role in maximizing the impact of VLSM research.

\section*{Acknowledgements}
We thank Yongsheng Zhang for sharing his Matlab implementation of the cluster size thersholding. We also thank Branch Coslett and the Laboratory for Cognition and Neural Stimulation and the Moss Rehabilitation Research Institute's Cognitive Area group for helpful discussions. We are particularly grateful to Dr. Myrna F. Schwartz and her research team for sharing the anatomical data that made these analyses possible and for her comments on an early draft of this report. This research was funded in part by Drexel University, University of Alabama at Birmingham, and National Institutes of Health grant R01DC010805 to DM.

\bibliography{ContinFWER}
\end{document}